\documentclass[setlength,byrevtex,aps,prl,groupedaddress,showpacs,floatfix,mathbbm,epsfig,graphics,twocolumn]{revtex4}

\usepackage{amsmath,amsfonts,amssymb,graphics,graphicx,epsfig,color,times,bbm,natbib,revsymb}

\def\l{\langle}
\def\r{\rangle}

\begin{document}

\title{Multi-Partite Entanglement Inequalities via Spin Vector Geometry}

\author{Gabriel A.\ Durkin$^{1}$} \email{gabriel.durkin@qubit.org}
\author{Christoph Simon$^{2}$}
\affiliation{ $^1$ Centre for Quantum Computation, Clarendon Laboratory, University of Oxford, OX1 3PU, U.K. \\
$^2$  Laboratoire de Spectrom\'{e}trie Physique, CNRS -
Universit\'{e} Grenoble 1, St. Martin d'H\`{e}res, France }

\date{9th April, 2005}

\pacs{03.65.Ud, 03.67.Mn}

\begin{abstract}
We introduce inequalities for multi-partite entanglement, derived
from the geometry of spin vectors. The criteria are constructed
iteratively from cross and dot products between the spins of
individual subsystems, each of which may have arbitrary
dimension. For qubit ensembles the maximum violation for our
inequalities is larger than that for the Mermin-Klyshko Bell
inequalities, and the maximally violating states are different
from Greenberger-Horne-Zeilinger states. Our inequalities are violated by certain bound
entangled states for which no Bell-type violation has yet been
found.
\end{abstract}

\maketitle

Entanglement is one of the most mysterious features of quantum
physics and a key ingredient in the science of quantum
information. While initial research was focussed on bipartite
entanglement, multipartite entanglement has attracted increasing
attention, because it was realized that multipartite entangled
states can exhibit qualitatively different features
\cite{GHZ-original}. Multi-partite entangled states are also
important for most applications envisaged in quantum information,
such as quantum computation \cite{NielsenandChuang}. Recently,
multi-partite entanglement has been studied using multi-partite
Bell inequalities \cite{mk,bell}, the partial transposition
criterion \cite{Dur-Bound-States}, and a variety of other methods
\cite{Acin;PRL02,other}.

Here we develop a new approach based on the geometry of spin
vectors. Our results are relevant for any system where operators
analogous to spin can be defined and measured, e.g. for multi-mode
light fields \cite{elaser} and Bose-Einstein condensates
\cite{Bose-Einstein-Spin-Crit}. In any finite-dimensional Hilbert
space one can always define operators $\{J_{x}, J_{y}, J_{z}\} =
\vec{J}$ that satisfy the commutation relations of angular
momentum, $[J_{u},J_{v}]= i \epsilon^{uvw}J_{w}$ for $u,v,w \in
\{x,y,z\}$. The transformations $U(\vec{\alpha}) = e^{i
\vec{\alpha} \cdot \vec{J}}$  generated by these operators, where
$\vec{\alpha}$ is a numerical vector, form a representation of the
group $SU(2)$. Applying  $U(\vec{\alpha})$ to the quantum states
leads to an $SO(3)$ rotation $R(\vec{\alpha})$ of the vector $\l
\vec{J} \r = \{\l J_{x} \r, \l J_{y} \r, \l J_{z} \r \}$ of spin expectation values. The vector $\vec{\alpha}$ gives both axis
and angle of rotation.

\emph{Dot Product}: Our inequalities involve the expectation
values of operators that are constructed from the spins of
subsystems via the cross and dot product. We will first illustrate
the principle with the simplest example, the dot product between
two spins of magnitude $j_1$ and $j_2$. For a product state
$|\psi_{12}\rangle=|\phi_1\rangle|\chi_2\rangle$ the expectation
value of the dot product is 
\begin{equation} \langle
\psi_{12}|\vec{J}^{(1)} \cdot
\vec{J}^{(2)}|\psi_{12}\rangle=\langle
\phi_{1}|\vec{J}^{(1)}|\phi_1\rangle \cdot \langle \chi_2
|\vec{J}^{(2)}|\chi_2\rangle,
\end{equation}
which is the scalar product of two vectors $\langle
\vec{J}^{(1)}\rangle=\langle \phi_{1}|\vec{J}^{(1)}|\phi_1\rangle$
and $\langle \vec{J}^{(2)}\rangle=\langle \chi_2
|\vec{J}^{(2)}|\chi_2\rangle$. The modulus of the scalar product
is bounded by the product of the norms of the two vectors.
Furthermore the norm $||\langle \vec{J}^{(1)}\rangle||$ cannot
exceed $j_1$. This can be seen by noting that by a rotation the
vector $\langle \vec{J}^{(1)}\rangle$ can always be brought to a
form where only one of its components, say $\langle J_z^{(1)}
\rangle$, is different from zero, without changing its norm. One
has $|\langle J_z^{(1)} \rangle| \leq j_1$, because $j_1$ and
$-j_1$ are the eigenvalues of $J_z^{(1)}$ with the largest
modulus. As a consequence, we have for every product state
\begin{equation}\label{new criterion2}
| \l D^{(2)} \r |=|\langle \vec{J}^{(1)} \cdot \vec{J}^{(2)}
\rangle| \: / (j_1 j_2) \leq 1.
\end{equation}
The bound of Eq. \eqref{new criterion2} can be
extended straightforwardly to separable states given the following
triangle inequality (the expectation value for product states has
suffix $\psi$):
\begin{align}
 \left| \sum_{\psi} \text{p}_{\psi}  \langle \vec{J}^{(1)} \cdot \vec{J}^{(2)} \rangle_{\psi}\right| \;   \leq  \; \sum_{\psi} \text{p}_{\psi} \left|  \langle \vec{J}^{(1)} \cdot \vec{J}^{(2)} \rangle_{\psi} \right|
\end{align}
Note that the choice of co-ordinate system for each spin is
arbitrary.

What is the maximal violation of Eq. \eqref{new criterion2} for
entangled states? Noting the relation $(\vec{J}^{(1)}.
\vec{J}^{(2)}) = \frac{1}{2} (\vec{J}^{\: 2} - \vec{J}^{(1) \: 2} -
\vec{J}^{(2) \: 2})$, with $\vec{J} =\vec{J}^{(1)} + \vec{J}^{(2)}$, indicates that the eigenstates of $(\vec{J}^{(1)}.
\vec{J}^{(2)})$ are also eigenstates of $\vec{J}^{\: 2}$,
corresponding to $j(j+1)$ where $j = j_1 + j_2 - \lambda$, and
$\lambda \in {\mathbbm N}_{0} \leq 2 \text{Min}[j_1,j_2] $.
Assuming $ j_1 \leq j_2$ and that subspaces $(1)$ and $(2)$ have
fixed dimension  gives:
\begin{align}
- \left( \: 1 + \frac{1}{j_2} \: \right) \leq \frac{1}{j_1 j_2}
\left(\l \vec{J}^{(1)}.  \: \vec{J}^{(2)} \r \right) \leq 1
\end{align}
From this result it is seen that the maximum absolute value is
$3$, provided by the $2$-qubit singlet, $j_1 = j_2 = 1/2$.

\emph{Cross Product and Multipartite Inequalities}: In full
analogy with the dot product, the expectation value of the cross
product in a product state is the cross product of the spin
expectation value vectors for individual systems,
\begin{equation}
\langle \psi_{12}|\vec{J}^{(2)} \times
\vec{J}^{(1)}|\psi_{12}\rangle=\langle
\chi_{2}|\vec{J}^{(2)}|\chi_2\rangle \times \langle \phi_1
|\vec{J}^{(1)}|\phi_1\rangle.
\end{equation}
As the norm of the cross product of two vectors is bounded from above by
the product of the their norms, we find for product
states
\begin{equation} \label{C2bound}
||\l \vec{C}^{(2)} \r||=||\langle \vec{J}^{(2)} \times
\vec{J}^{(1)} \rangle||  \: / \: (j_1 j_2)  \;  \leq 1,
\end{equation}
and again the generalization to separable states is immediate.

By iterating the cross product operation it is possible to derive
bounds for multipartite systems. For a fully separable state of
$N$ spins $j_1, j_2, j_3, ... , j_N$ one derives the
following bound, in analogy with Eq. \eqref{C2bound}:
\begin{equation} \label{Cross-ineq}
||\langle \vec{C}^{(N)} \rangle|| \leq 1
\end{equation}
where we use the notation:
\begin{equation}\label{multi-crit-vec}
\vec{C}^{(N)}=\vec{J}^{(N)} \times (\vec{J}^{(N-1)} \times ...
(\vec{J}^{(2)} \times J^{(1)}))/(j_1 j_2 ... j_N).
\end{equation}
Considering the dot product between a single spin and a vector
constructed like in Eq. \eqref{multi-crit-vec} one can also derive
a bound 
\begin{equation}\label{Dot-ineq}
|\langle D^{(N)}\rangle| \leq 1.
\end{equation} 
for fully separable states, where
\begin{equation}\label{multi-crit-scal}
D^{(N)}=\vec{J}^{(N)} \cdot (\vec{J}^{(N-1)}... \times
(\vec{J}^{(2)} \times J^{(1)}))/(j_1 j_2 ... j_N).
\end{equation}
We have investigated how strongly the bounds of Eq. \eqref{Cross-ineq} and Eq. \eqref{Dot-ineq} can be violated by entangled states, TABLE I. For $D^{(N)}$ this involves
finding its largest eigenvalue. Maximum values of
$||\vec{C}^{(N)}||$ may be found by studying the greatest
expectation value for any of the vector components of
$\vec{C}^{(N)}$. This is because joint identical SU(2)
transformations on all spins (i.e. each with the same
$\vec{\alpha}$) correspond to simple rotations of the vector
$\langle \vec{C}^{(N)}\rangle$. It can thus always be brought to a
standard form where only one of its components e.g. $C^{(N)}_z$ is
non-zero, without changing its norm. The maximum of the norm is
therefore the largest eigenvalue of $C^{(N)}_z$. Upper bounds also
exist for all entanglement partitions and one partition may have a
range of upper bounds depending on the ordering of sub-systems
$(1),(2),\ldots, (N)$ in the directed products $D^{(N)}$ and
$\vec{C}^{(N)}$, see TABLE II.

\begin{table}\center
\nonumber
\begin{tabular}{|c||c|c|}
\hline  \hline
$ \: (N,J) \:$  & $ \; \text{Max}\left|\left|  \l \vec{C}^{(N)} \r  \right|\right|  \;  $ & $\text{Max}\left|  \l D^{(N)} \r  \right|  $ \\
\hline
\: $(2,1/2)$ \: & $2 $ & $3$ \\
 \: $(3,1/2)$ \: & $2 \sqrt{2} \; (\approx 2.828) $ & $2 \sqrt{3} \; (\approx 3.464)$ \\
\: $(4,1/2)$ \: &  $2 \sqrt{6} \; (\approx 4.899)$ & $4 \sqrt{3} \; (\approx 6.928)$  \\
\: $(5,1/2)$ \: & $2\sqrt{14}  \; (\approx 7.483)$ & $4 \sqrt{6} \; (\approx 9.798)$  \\
\: $(6,1/2)$ \: &$\approx 12.144$  &$\approx 16.971$ \\
 \hline
 \: $(2,1)$&  $\sqrt{2} \; (\approx 1.414) $ & $2$\\
\: $(3,1)$   &  $\sqrt{3} \; (\approx 1.7321) $ & $\sqrt{3} \; (\approx 1.7321)$\\
\: $(4,1)$ & $\sqrt{3+\sqrt{5}} \; (\approx 2.288)$  & $2\sqrt{2} \; (\approx 2.828)$   \\
\: $(5,1)$&  $\approx 2.840$ &  3 \\
\: $(6,1)$ &$ \approx 3.731$   &  $\approx 4.472$ \\
 \hline \hline
\end{tabular}
    \caption{\small{Some maximal violations for the inequalities of Eq. \eqref{Cross-ineq} and Eq. \eqref{Dot-ineq},  for entangled states of $N$ qubits $(J=1/2)$ and qutrits $(J=1)$. The largest violation we were able to find numerically was $\text{Max}\l D^{(11)} \r \approx 152.691$, for $11$ qubits. $D^{(11)}$ has $2^{11}$ eigenvalues with some degeneracy, see Eq. \eqref{Dstructure}. }}
 \end{table}

\begin{table}[b]\center 
\nonumber
\begin{tabular}{|c|c|c|c|}
\hline  \hline
 \; Partition  \; & $ \; \; \text{Max}||  \l \vec{C}^{(4)} \r || \; \; $ & $\text{Max}|  \l D^{(4)} \r  | $& $ \; \; \text{Max} \l F^{(4)} \r \; \;$  \\
\hline
$[1 \: 2 \: 3\: 4]$& $2\sqrt{6} \: (\approx 4.899) $ & $4 \sqrt{3} \: (\approx 6.928)$ & $2\sqrt{2} \: (\approx 2.828) $\\
$[1\: 2\: 3|4]$  & $2\sqrt{2} \: (\approx 2.828)$ & $ 2\sqrt{2}  $  &  $2 $\\
$[1\: 2|3 \: 4]$ & $4$ & $6$  & $\sqrt{2}$ \\
$[1|2|3\: 4]$& $2$ & $2$ & $\sqrt{2} \: (\approx 1.414)$\\
$ [1|2|3|4]$ &$1$  &$1$ &$ 1$\\
 \hline \hline
\end{tabular}
    \caption{\small{Magnitudes of $ \l \vec{C}^{(4)} \r $ and $ \l D^{(4)} \r $  have distinct upper bounds (found numerically) for entanglement partitions of 4 qubits, labelled `$(1)$' to `$(4)$'. No entanglement exists between qubits separated by a vertical bar. There are $4!$ permutations for the directed products $\vec{C}^{(4)}$ and $ D^{(4)}$, resulting in a range of upper bounds for some partitions, the largest of which is shown above. For example, $[1|2|3\: 4]$ has three distinct bounds for $\text{Max}||  \l \vec{C}^{(4)} \r ||$, namely $1$, $\sqrt{2}$ and $2$. Note that the Mermin-Klyshko operator $ F^{(4)} $ gives identical upper bounds \cite{Gisin-MK} for $[1|2|34]$ and $[12|34]$; its expectation value is also degenerate under all particle re-orderings. The new inequalities produce a larger maximal violation. Because the new partition bounds are the result of a (global) numerical optimization, greater computing power and a more refined search may allow some of them to be improved.}}
 \end{table}

\emph{Symmetry and Eigenstates:} In addition to the numerical
results, some insight into the structure of the eigenstates
may be gained from symmetry considerations. $\l D^{(N)} \r $ and
$ \l \vec{C}^{(N)}\r $ transform like a scalar and a vector
respectively under all identical joint rotations
$R(\vec{\alpha})^{\otimes N}$, or equivalently under all identical
local $SU(2)$ transformations of the state. Furthermore, $D^{(N)}$
is anti-symmetric under the permutation $(1)\leftrightarrow(2)$
because $(\vec{J}^{(1)} \times \vec{J}^{(2)}) = -(\vec{J}^{(2)}
\times \vec{J}^{(1)})$. Operator $D^{(N)}$ must therefore have the
following highly symmetric structure:
\begin{equation}\label{Dstructure}
D^{(N)} = \sum_{{\mathcal D}} \mu_{{\mathcal D}}
(\Pi^{(12)}_{{\mathcal D}}- \Pi^{(21)}_{{\mathcal D}})
\end{equation}
i.e. a weighted sum of projectors $\Pi_{{\mathcal D}}$ onto spaces
associated with irreducible matrix representations ${\mathcal D}$
of $SU(2)$. All the projectors are orthogonal, $\Pi_{{\mathcal D}}
\Pi_{{\mathcal D}'} = \delta_{{\mathcal D} {\mathcal D}'}
\Pi_{{\mathcal D}}$ and they each project into a $(2J+1)$
dimensional space of overall spin $J({\mathcal D})$. 
States in each representation ${\mathcal D}$ have a shared permutation symmetry
of all the $N$ particles; e.g. the highest spin representation
${\mathcal D}^{*}$, for which $J=j^{(1)} + \ldots + j^{(N)}$, is
inclusive of all states symmetric under all particle permutations.
The projector $\Pi^{(21)}$ is formed from $\Pi^{(12)}$ by
exchanging particles labelled `$(1)$' and `$(2)$', mapping either
$\Pi^{(12)}_{{\mathcal D}}$  to itself, in which case it vanishes
from Eq. \eqref{Dstructure}, or to an orthogonal projector of the
same spin. Thus, eigenvalues of $D^{(N)}$
appear in pairs of opposite sign $\pm(\mu_{{\mathcal D}}-\mu_{{\mathcal D}'})$, with the multiplicity of
pairs $2J({\mathcal D})+1$. To give a concrete example,  eigenstate $|\psi_4 \r$ maximally
violates $| \l D^{(4)} \r | \leq 1$:
\begin{align}\label{4qubitsinglet}
\! \!\! \!\! \!\! \!\! \! \! \!\! \!\! \!\! \!\! \!\! \! 2 \sqrt{6} \: |\psi_4 \r = (1+ \sqrt{3}) (|\uparrow \uparrow \downarrow \downarrow \r + |\downarrow \downarrow \uparrow \uparrow \r)   \nonumber \\ \!\! \!\! \!\! \!
+(1- \sqrt{3})(| \downarrow  \uparrow \downarrow \uparrow \r +   | \uparrow \downarrow \uparrow \downarrow \r) -2 (|\downarrow
\uparrow \uparrow \downarrow \r + |\uparrow \downarrow \downarrow
\uparrow  \r )
\end{align}
This state is one of two orthogonal $J=0$ states of $4$ qubits,
and gives $D^{(4)} |\psi_4 \r =  4 \sqrt{3} |\psi_4 \r$.

Since $\vec{C}^{(N)}$ is a vector
operator  i.e. $[C^{(N)}_{u},J_{v}]= i
\epsilon^{uvw}C^{(N)}_{w}$ for $u,v,w \in \{x,y,z\}$, it is a spin-1 object, of which $C^{(N)}_{z}$ is the $m=0$
component. As a consequence and in contrast to $D^{(N)}$, operator $C^{(N)}_{z}$ is not diagonal in the spin basis. The Wigner-Eckart theorem
\cite{Sakurai-Wigner} may be invoked to reveal that
\begin{equation}\label{Cstruct-WigEck}
 \l J',m' |C^{(N)}_{z}| J, m \r  =  \l J,1;m,0|J,1;J',m' \r \: T_{J,J'}
\end{equation}
where $\l j_1,j_2;m_1,m_2 |j_1,j_2;j,m\r$ are the Clebsch-Gordan
coefficients and $T_{J,J'}$ is a transition matrix element
dependent only on $J$ and $J'$. The spin-1 selection rules are the familiar ones of dipole radiation; $(J'-J)$ and $(m'-m) \in \{-1,0,1\}$, with the $J=J'=0$ transition forbidden, i.e. $ T_{0,0}=0$. From this perspective, the scalar operator $D^{(N)}$ is a spin-$0$ object, only able to couple states for $J=J'$ and $m=m'$.  

The
pair anti-symmetry of $D^{(N)}$ is also true of $C^{(N)}_{z}$ (eigenvalues appear in
pairs $\pm \chi$). An example of the states maximally violating
Eq. \eqref{Cross-ineq} is $|\phi_{4} \r$, 
\begin{align}\label{4qubitmaxC}
4 \sqrt{3} \: |\phi_4 \r = 3 (|\uparrow \downarrow \uparrow \uparrow \r - |\downarrow \uparrow \downarrow \downarrow \r)   \nonumber \\
+\sqrt{6}(|   \uparrow \downarrow \downarrow \uparrow \r +   |
\downarrow \uparrow  \uparrow \downarrow \r
- |   \uparrow \downarrow  \uparrow \downarrow \r -  | \downarrow \uparrow \downarrow \uparrow  \r) \nonumber \\
+ |\downarrow  \downarrow  \downarrow \uparrow \r - |\uparrow
\uparrow \uparrow \downarrow \r + |\downarrow  \downarrow \uparrow
\downarrow  \r - |\uparrow \uparrow \downarrow \uparrow  \r + |
\uparrow \downarrow  \downarrow \downarrow  \r - | \downarrow
\uparrow \uparrow  \uparrow  \r
\end{align}
which gives $C^{(4)}_{z} |\phi_{4} \r = 2 \sqrt{6} |\phi_{4} \r $,
cf. TABLE II. This is one of four orthogonal states that give the
same maximum.

\emph{Comparison with Mermin-Klyshko Inequalities:} There are
$2^{2^{N}}$ independent Bell inequalities for $N$ qubit ensembles
having two two-valued observables per qubit or
`site', \cite{Werner-Wolf-22NBell}. This set is the simplest and
best understood of multi-partite Bell inequalities, although
others may be formulated with e.g. $3$ observables per site,
\cite{three-setting-bell-to-detect-Dur}. The $2^{2^{N}}$
inequalities are satisfied by all local hidden variable theories,
and all are maximally violated by GHZ states:
\begin{equation}
| \text{GHZ}_{N} \r =( \: |  \uparrow \r^{\otimes N} + |
\downarrow \r^{\otimes N} \: ) / \sqrt{2}.
\end{equation} 
Note that the
states which maximally violate Eq. \eqref{Cross-ineq} and Eq.
\eqref{Dot-ineq}, e.g. $| \phi_4 \r$ and $| \psi_4 \r$,
are generally inequivalent to $| \text{GHZ}_{4} \r$ under local unitaries.
This can be proved by determining the Schmidt
coefficients of the states for a bipartite $\{2,2\}$ cut.

\begin{figure}
\includegraphics[width=3.2in]{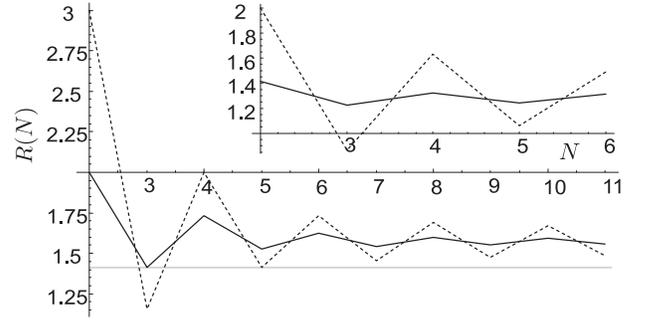}
\caption{\small{Ratio of successive maximal violations for qubits
and (inset) qutrits. $R(N) \mapsto \text{Max}||\l \vec{C}^{(N)} \r
||  / \text{Max}|| \l \vec{C}^{(N-1)} \r ||$  (unbroken line), and
$ \mapsto \text{Max}| \l D^{(N)} \r |  /  \text{Max}| \l D^{(N-1)}
\r |$ (dashed line). $D^{(1)}$ is unity, and $\vec{C}^{(1)} =
\vec{J}/j$. The
ratio for Mermin-Klyshko inequalities is $\sqrt{2}$ (grey
horizontal line). }} \label{crossanddot}
\end{figure}

For the Bell inequalities of the type mentioned, the
Mermin-Klyshko (MK) inequality \cite{mk} has the largest possible
violation \cite{Werner-Wolf-22NBell}, within the context of
quantum mechanics. The MK inequality depends on operator functions
$F^{(N)}$ that may be defined recursively:
\begin{equation}
2 F^{(N)} \! = F^{(N-1)} \! \otimes \! (A^{(N)} \! + \!
\tilde{A}^{(N)}) +  \tilde{F}^{(N-1)} \! \otimes \! (A^{(N)} \! -
\! \tilde{A}^{(N)})
\end{equation}
where $A^{(N)}$ and $\tilde{A}^{(N)}$ are observables of the $N$th
qubit with eigenvalues $\pm 1$, and $F^{(1)} = A^{(1)}$. Operator
$\tilde{F}^{(N)}$ is obtained under exchange $A \leftrightarrow
\tilde{A}$ for all the nested observables. The MK inequality
satisfied by local hidden variable theories is $\l F^{(N)} \r \leq
1$.

For any quantum state, $F^{(N)}$ has an upper bound, $\l F^{(N)} \r \leq 2^{(N-1)/2}$, resulting in a smaller possible violation of the MK inequality above than that attainable for Eq.
\eqref{Cross-ineq} and Eq. \eqref{Dot-ineq} in qubit ensembles $N
\mapsto \{2,3,4,\ldots,11\}$, cf. TABLE I. Also, the ratio of maximal violation
for $N$ qubits to $(N-1)$ qubits is always $\sqrt{2}$ for the MK
inequalities, whereas for the inequalities of Eq.
\eqref{Cross-ineq} the ratio is always $\geq \sqrt{2}$ but
displays an intriguing `see-saw' character; FIG.\ref{crossanddot} gives ratios
for qubits and qutrits, and for both Dot and Cross inequalities.
That a larger violation is possible for the new inequalities may
be attributable at least in part to having three (orthogonal)
observables per qubit, compared to two for the MK inequalities.
Another possible reason is that our inequalities do not exclude
the possibility of a local hidden variable model. They are
strictly criteria for non-separability.

Operators $\vec{C}^{(N)}$ and $D^{(N)}$ are also unlike $F^{(N)}$ in that they are not symmetric under particle
exchange; $\vec{C}^{(N)}$ is a `directed' product. TABLE II shows
how different particle orderings results in different bounds on
$\vec{C}^{(4)}$ and $D^{(4)}$.

In terms of the numerical search needed to find a violation, three
parameters will define $3$ orthogonal directions in ${\mathbbm
R}^{3}$, whereas $4$ variables are needed for both measurement
directions per subspace in a Bell inequality. Thus, fewer
parameters are required in the optimisation of Eq.
\eqref{multi-crit-vec} and Eq. \eqref{multi-crit-scal}.

\emph{Entanglement detection and Robustness:} Consider
$D^{(3)} = \vec{J}^{(3)} .( \vec{J}^{(2)} \times \vec{J}^{(1)}) /
j_1  j_2  j_3 $ for $3$ qubits. Its largest amplitude
eigenvalue is $2\sqrt{3}$, an associated eigenstate is
\begin{equation}\label{W-state}
|W_3 \r  =  ( \: | \uparrow \uparrow \downarrow \r + e^{i \alpha}|
\uparrow \downarrow \uparrow \r + e^{i \beta} | \downarrow
\uparrow \uparrow \r\:)/\sqrt{3} \: ,
\end{equation}
a state inequivalent to $| \text{GHZ}_3 \r$ under rotations of the
local coordinate systems. For GHZ states, numerical optimisation
over all local coordinate systems gives $|\l D^{(3)} \r| \leq
\frac{3}{2} \sqrt{3}$, i.e. the maximum possible violation is
smaller than that for the $W$ state by a factor of $3/4$. For
completely separable states,  $|\l D^{(3)} \r|$ corresponds to the
volume of a parallelopiped with sides of unit length. The
detection of  $W$-type entanglement is robust against noise: Mixed
with white noise, $ (1- \nu) | W_3 \r \l W_3 | + \nu \: {\mathbbm
I}_{8}/8$, the fraction of noise $\nu$ can be as high as $71 \%$
and $|\l D^{(3)} \r| \leq 1$ is still violated. Substituting $|
\text{GHZ}_3 \r$, violation occurs for $\nu \leq 61 \%$.
The $3$ qubit MK inequality will only detect $| \text{GHZ}_3 \r$
mixed with less than $50\%$ ${\mathbbm I}_{8}/8$, even though it
is maximally violated by such states.

\emph{Violation Ratio:} We now show that for a given entangled
state the maximum possible violation for the Cross criterion, when
optimizing the choice of local coordinate systems, cannot be
larger than the maximum violation for the Dot criterion. Consider
the correlation $i \in \{x,y,z \}$ defined as $i \equiv \l
J^{(N)}_{i} \otimes C^{(N-1)}_{i}\r /j^{(N)}$. One
may write:
\begin{equation}\label{DasXYZ}
D^{(N)} \! =  \! \! \! \sum_{i\in \{x,y,z\} } (J^{(N)}_i \otimes
C^{(N-1)}_i ) \left/ j^{(N)} \right. = x + y + z
\end{equation}
A correlation set $\{x,y,z\}$ may be mapped into $ \{x,-y,-z\},
\{-x,y,-z\}$ and $\{-x,-y,z\}$ by local unitaries ($\pi$ rotations
of the $N$th qubit about $x,y,z$ axes respectively). Therefore
$\text{Max}|\l D^{(N)} \r|$ is associated with correlations $ \l
J^{(N)}_i \otimes C^{(N-1)}_i \r$  all having the \emph{same}
sign, i.e. $\text{Max}|\l D^{(N)} \r| = \text{Max}(|x|+|y|+|z|)$.
For the Cross criterion, taking $\text{Max}||
\l\vec{C}^{(N)}\r|| = \text{Max}| \l C^{(N)}_{z} \r|$, one has:

\begin{align}\label{Cz-mapsto}
&\text{Max}| \l C^{(N)}_{z} \r| \! = \!   \text{Max}\left | \left
\l \frac{J^{(N)}_{y}}{j^{(N)}} \! \otimes \! C^{(N-1)}_{x} \! - \!
\frac{J^{(N)}_{x}}{j^{(N)}} \! \otimes  \! C^{(N-1)}_{y} \right \r
\right| \nonumber \\ &=  \! \left.  \text{Max}\left| \left\l
J^{(N)}_{x}  \! \otimes \! C^{(N-1)}_{x} \! + \!  J^{(N)}_{y}  \!
\otimes  \! C^{(N-1)}_{y} \right \r \right | \: \right/ j^{(N)}
\end{align}
(which is $\text{Max}|x+y|$) because a
local rotation in the $N$th subspace transforms
$\{J^{(N)}_{x},J^{(N)}_{y} \} \mapsto \{-J^{(N)}_{y},
J^{(N)}_{x}\}$ above. Thus for a given state, the maximum of $| \l
C^{(N)}_{z} \r|$  for all choices of local coordinate systems is
at most a sum of two of the absolute values, $|x|$, $|y|$ and
$|z|$, obviously upper-bounded by $\text{Max}(|x|+|y|+|z|)$
\begin{equation} \label{ratio}
\text{Max}|| \l \vec{C}^{(N)} \r || \leq \text{Max}| \l D^{(N)} \r
| \end{equation}
It is stressed that for Eq. \eqref{ratio} the particles or
subspaces are considered in the same order for both
$\vec{C}^{(N)}$ and  $D^{(N)}$.

\emph{Bound Entanglement:}  We consider a mixture of GHZ and
product state projectors \cite{Dur-Bound-States}:
\begin{align}
\! \! \rho_N \! = \! \frac{1}{N \! + \! 1} \! \left( \!
|\text{GHZ}_{N} \r \l \text{GHZ}_{N} | + \frac{1}{2}
\sum_{n=1}^{N}(\Pi_{n} + \Pi_{\tilde{n}}) \right)
\end{align}
Here $\Pi_{n}$ is a projector onto product state \linebreak
$|\uparrow \r_{1} |\uparrow \r_{2} \ldots |\downarrow \r_{n}
\ldots |\uparrow \r_{N}$, i.e. only the $n$th qubit is in the
orthogonal state. Projector $\Pi_{\tilde{n}}$ is obtained from
$\Pi_{n}$ by interchanging all $\uparrow$ with $\downarrow$.
In \cite{Dur-Bound-States} it was shown that 
$\rho_N$ is entangled for $N \geq 4$ (it has negative partial
transposition in some partitions) but entanglement cannot be
distilled for any $\{1,(N-1)\}$ partition, it is `bound'
\cite{pereshorodecki3}. See also \cite{Acin;PRL02}. The state
violates $\l F^{N} \r \leq 1$ if and only if $N \geq 8$ and
violates a Bell inequality of three dichotomic observables per
qubit \cite{three-setting-bell-to-detect-Dur} for $N \geq 7$;
recently it was shown to violate a `functional' Bell inequality
\cite{N6Bell} for $N \geq 6$. Bell violations for lower $N$ have
yet to be shown. In contrast, the $N=4$ state violates both
`Cross' and `Dot' inequalities: A numerical search over local
unitaries gives $\text{Max}|| \l \vec{C}^{(4)} \r || \approx 1.09$
and $ \text{Max} |\l D^{(4)} \r | \approx 1.25$ for $\rho_4$. This
result was unexpected, compared with the non-violation of MK
inequalities by $\rho_4$. After all, the only entanglement in
$\rho_N$ is due to the GHZ state -- this maximally violates the MK
inequality.

\emph{Summary:} A simple geometric approach allows the
formulation of entanglement inequalities based on the scalar and
vector products of two spin operators. This idea was extended to
inequalities for multiple subsystems, each of arbitrary
dimensionality. The maximum violation for $N$ qubits is greater
than that for certain classes of Bell inequalities, including the
Mermin-Klyshko inequalities. The maximally violating states are
not generally GHZ states; examples were given and elements of
their structure discussed. $D^{(3)}$ showed a high level of
robustness in detecting both $W$ and GHZ entanglement. Maximal
violations for all entanglement partitions of $4$ qubits were
found numerically -- none of them is degenerate (unlike the MK
bounds), and maximum violation is for a fully $4$-entangled state.
We have also shown that the new inequalities can detect bound
entanglement in states for which a Bell violation has not been
found. The work here followed a spin description, but may be
applied to arbitrary systems for which $\{J_x,J_y, J_z\}$
operators can be defined and measured.

\begin{acknowledgments}
We would like to thank D. Bouwmeester for useful comments.
\end{acknowledgments}

\bibliographystyle{apsrev}

\end{document}